Article type (Review)

# Replicated Spectrographs in Astronomy

Gary J. Hill

Hill@astro.as.utexas.edu
McDonald Observatory and Department of Astronomy, University of Texas at Austin, 2515 Speedway, Stop C1402, Austin, TX 78712-1205, USA

**Abstract**
As telescope apertures increase, the challenge of scaling spectrographic astronomical instruments becomes acute. The next generation of extremely large telescopes (ELTs) strain the availability of glass blanks for optics and engineering to provide sufficient mechanical stability. While breaking the relationship between telescope diameter and instrument pupil size by adaptive optics is a clear path for small fields of view, survey instruments exploiting multiplex advantages will be pressed to find cost-effective solutions.

In this review we argue that exploiting the full potential of ELTs will require the barrier of the cost and engineering difficulty of monolithic instruments to be broken by the use of large-scale replication of spectrographs. The first steps in this direction have already been taken with the soon to be commissioned MUSE and VIRUS instruments for the Very Large Telescope and the Hobby-Eberly Telescope, respectively. MUSE employs 24 spectrograph channels, while VIRUS has 150 channels. We compare the information gathering power of these replicated instruments with the present state of the art in more traditional spectrographs, and with instruments under development for ELTs.

Design principles for replication are explored along with lessons learned, and we look forward to future technologies that could make massively-replicated instruments even more compelling.

1. Introduction

Since Fraunhofer first combined a prism with a telescope to observe spectra of planets and bright stars in the early 19$^{th}$ Century, the traditional optical or near infrared (NIR) astronomical instrument has had a monolithic design and is a one-off prototype, where a large fraction of the cost is expended on engineering effort. As telescope size increases, geometric considerations are forcing us to the point where the size and cost of spectrographs for the current generation of very large telescopes (VLTs, 8-10 m class) is approaching a limit. The physical size of VLT instruments (particularly those mounted at Nasmyth foci) is such that careful design and active correction of flexure is needed to maintain instrument alignment. This complexity is reflected in the $20M price tags for the most capable instruments, when engineering costs are fully accounted.

Large-scale replication of spectrographs, combined with various modes of slicing the information gathered by large telescopes, promises to break the relation between telescope aperture growth and instrument size and cost. The first steps into this regime on VLTs have been made with the Multi Unit Spectroscopic Explorer (MUSE [1]) for the ESO Very Large Telescope (VLT) and the Visible Integral-field Replicable Unit Spectrograph (VIRUS [2]) for the Hobby Eberly Telescope (HET).

In this review we examine the information-gathering power of current astronomical spectrographs and conclude that for important areas of astronomical research the next wave of instruments must press into the replicated regime. We examine the advantages of this approach, enabling technologies, applications to extremely large telescopes (ELTs, 25-40 m class), and future directions. This review is intended for non-experts in astronomical instrumentation and aims to summarize the current state of the thrust into spectrograph replication epitomized by the first two highly-replicated VLT instruments (MUSE and VIRUS) entering use in the near future and the LAMOST facility that has started operations [3]. We will also examine the drivers to replication provided by the next generation of ELTs. The focus is on low to moderate spectral resolution instruments since these are most often employed for large-area surveys where maximizing the number of objects observed simultaneously is an important science driver. Many telescopes and instruments will be mentioned and rather than spell out all acronyms, the reader is referred to the respective publications.

## 2. The drive to replication

As the size of telescopes increases, driven by the need to gather more light and provide greater spatial resolution, in combination with adaptive optics (AO [4]) techniques, the physical size of spectrographs increases. For a spectrograph, as telescope diameter D increases at fixed focal ratio F, the size of the collimator and pupil also increase in unison, and the camera follows. The plate scale ($\propto (DF)^{-1}$ in arcseconds/mm) of the telescope also decreases linearly, so the number of pixels on the detector per spatial resolution element increases. As the resolution element becomes more over-sampled, the CCD area is less efficiently utilized and the amount of information gathered goes down as $1/D^2$. As a result, the overall angular field of the instrument on sky goes down and the number of spectral elements decreases. It is hard and expensive to make cameras faster and detectors bigger, and these geometric drivers become quite dominant for spectrographs on VLTs and ELTs.

With the current class of VLTs, the quest for a maximal combination of etendue (or areal grasp, telescope collecting area A, multiplied by angular area on sky sampled by the instrument Ω), spectral resolution, and wavelength coverage has led to physically large instruments with the largest detector areas possible. In several cases the instruments have been broken into 2 or 4 channels to increase the angular field of view to make surveys of large samples of objects more efficient. These instruments represent the first steps toward using replication of many copies of a base spectrograph to break the limitations of size, complexity, and cost associated with instruments for the largest telescopes. With 2-4 copies of the spectrograph channel, they do not realize the full cost-benefit of replication, since the engineering challenges and costs are still very high.

Adaptive optics [4] promises to alleviate this situation to some extent by enabling near-diffraction limited images over small fields, or improved images over larger fields, and greatly helps traditional spectrographs to be mated to the next generation of ELTs, without pushing beyond feasible sizes and costs. Even so, for surveys, instruments with very large refractive optics are being proposed, which limit the available glass choices needed to optimize the design and push the current state of the art, and which will be very expensive [5,6]. AO is not, however, the silver bullet that will prevent the cost and complexity of instruments from escalating, because:

- Extreme AO and Multi-Conjugate AO, where the Rayleigh limit is approached, is effective in the NIR over small fields of view, but practically ineffective at UV and blue wavelengths.
- Ground Layer AO (GLAO) offers the prospect of wider fields and improvements in the optical, but large instruments are still unavoidable.
- AO may keep the scale size of spectrographs observing single, or a few, objects within a controllable regime, but the number of spatial resolution elements within an interesting field of view is multiplied greatly. Access to the full information content potentially provided by a near-diffraction-limited ELT will require a new instrument paradigm.

Since the physical size of the diffraction limit is independent of aperture at a given focal ratio, the drive to AO on ELTs ensures that instruments do not have to grow as fast as aperture to couple efficiently to the telescope. However, the angular area covered by these instruments still goes down as $1/D^2$, and with it etendue or areal grasp. If the full field of view of ELTs is to be exploited, there is still impetus to push towards replicating instruments. For reference, a 30 m with an extreme AO field of view of 1 arcminute (or with a GLAO field of 10 arcminutes) has as many resolved spatial elements (50 million) as a 4 m with a 2 degree field on a site with decent natural seeing. With the large light-gathering power of ELTs, the surface density of suitable targets on sky is at least as high as on the 4 m.

Grasp is a useful measure of the performance of a spectrograph, since the quantity is independent of aperture, so long as the input focal ratio to the spectrograph is maintained. On small telescopes, the spatial element may get too large in projection to fulfil the science case, and on the largest telescopes it may get too small to capture the light from a single object. Slicing techniques (Section 3) can aid in these regimes, but large grasp is an essential feature of spectrographs deployed for surveys.

In Figure 1, following Bershady [7,8], we plot instrument areal grasp (AΩ) versus spectral power (the product of the number of spectral resolution elements and the resolving power $N_\lambda R$, where $R=\lambda/\delta\lambda$ for wavelength $\lambda$ and spectral resolution $\delta\lambda$) as a useful metric in following the evolution of total instrument grasp (TIG, the product of the two quantities). In Figure 1 we show the locus of these quantities for different-sized detectors, extending up to the maximum current CCD pixel-count ($N_{pxl}$) used in spectrographs of about 8kx8k pixels (or 6kx12k) achieved on single wafers or small mosaics. The loci for each detector size are plotted for the following instrument parameters: f/1.3 camera focal ratio (a practical limit for refractive optics), 15 μm pixel size typical of modern CCDs used in astronomy, 4.0 pixels per resolution element with a packing efficiency of 65% to allow for separation of spectra on the detector, and resolving power of R=1000 typical of large survey spectrographs on a range of telescope sizes. Anamorphic magnification effects of the disperser are ignored.

The loci at constant $N_{pxl}$ are diagonal and form parallel lines with increasing $N_{pxl}$ or TIG. The loci reflect the fact that the available detector area is divided between spatial and spectral elements depending on details of instrument design driven by science requirements. Available detector size is a primary driver of the evolution of spectrograph capability. The evolution in CCD format and opto-mechanical engineering methods means that the progression towards higher TIG has been fairly monotonic with time. This progression is accompanied by an inherent increase in instrument size and cost. The 8kx8k locus can be considered a practical upper limit for monolithic instruments. While larger formats can be built up with mosaics, the instruments quickly press the limits of refractive optical element size and mechanical engineering constraints, and it is not an option to make faster cameras in order to utilize the pixels more efficiently.

Integral field spectrographs (IFS, which divide a 2D field of view into spatial elements that are dispersed and recorded simultaneously, hence creating what is referred to as 3D or integral-field spectroscopy) and multi-object spectrographs (MOS) for surveys are the instruments that drive to high grasp/etendue, currently. Both drive to gather spectra from the largest possible number of spatial elements at once. In Figure 1 we plot a compendium of existing IFS compiled by Bershady [8] and several major MOS (Table 1) to illustrate the current state of the art. The points have an upper bound close to the boundary associated with the largest detectors, with some displaced above the line due to their focus on higher spectral resolution. A number of instruments are highlighted. We include WEAVE, a 1000-fiber MOS with R~5,500 to 25,000 under development for the 4 m WHT [9] and 4MOST proposed for the 4 m VISTA [10]. In addition, the HERMES instrument for AAT stands out due to its emphasis on much higher resolution for galactic archaeology [11]. These three instruments all push the boundaries of detector size for single spectrographs and employ dichroic beamsplitters to separate and simultaneously disperse and detect light from multiple spectral ranges (see next Section).

The shaded region of higher TIG beyond the 8kx8k $N_{pxl}$ locus in Figure 1 is the domain of replicated instruments. The 4 m instruments HERMES, WEAVE and 4MOST are able to use the CCD pixels more efficiently than assumed for the detector size loci in the figure, due to the coarser plate-scale of these smaller telescopes, and hence are able to rise above this locus, but could not be adapted to larger telescopes without losing light. In Table 2 we gather information about all replicated spectrographs in use or under development, and plot their total grasp and spectral power in Figure 1. We define replicated as any instrument with more than one identical spectral channel. In reality the cost and engineering advantages of using replication to increase TIG are not really realised until 10s of copies are made (large-scale replication), and we define "massive" replication as over 100 identical channels. However, the instruments with 2-4 channels, starting with the Palomar 4-shooter instrument [12], have provided impetus for the replication concept and deserve recognition as such. The more highly-replicated instruments MUSE, LAMOST, and the massively replicated VIRUS are highlighted in Figure 1.

The areal total grasp plotted in the figures is a surrogate for etendue, which in astronomy terminology also includes the efficiency of the instrument, and which would be the ideal quantity to plot, being most directly related to the number of photons detected. However, it is hard to establish consistent measures of efficiency between instruments, and since most are designed to have the highest possible efficiency within limitations of grating and detector efficiency, and quality of optics coatings, grasp is a good surrogate. Note that increasing spectral resolution will formally move instruments to the right in Figure 1 at fixed $N_{pxl}$, increasing spectral power at the expense of spectral coverage, which is not captured in this diagram. There are limits where it becomes necessary to cross-disperse the spectra, converting CCD real-estate to spectral coverage at the expense of areal grasp. These limits are in the R~10-20k regime. The other effect of increasing R, not captured in the diagram is the reduction in efficiency of dispersing elements, which would decrease etendue but not areal grasp. The focus of this review is on moderate resolution survey instruments, so these factors are side-stepped here.

## 3. Multiplexing modes

Instrumental pupil diameter growth can be mitigated by multiplexing. Physical limitations on the size of detectors and optics mean that in order to increase etendue/grasp or spectral coverage/resolution various multiplexing or slicing methods are employed. Slicing can be used to increase the grasp or the spectral coverage and resolution, depending on the science drivers.

There are several ways to approach multiplexing, all of which require more detector area, but allow the scale of individual spectrograph channels to be controlled:
- Field division into multiple spectrographs
- Spectral slicing within spectrographs
- Image slicing of a single image or small field of view into a single spectrograph
- Pupil slicing of a single image into single or multiple spectrographs

Field division has become a favoured choice for large imaging spectrographs. The simplest form is to array two or more spectrographic channels viewing adjacent areas of sky within the field of view of the telescope. Examples include DEIMOS, and VIMOS. These instruments are classified as replicated in Table 1 and Figure 1, but do not achieve significant economies of scale, partly because of the engineering challenges of keeping the registration of the multiple fields aligned and of the sheer scale of the instruments. MUSE and VIRUS employ a high degree of field division (24- and 75-fold respectively) as the first step in achieving their huge grasp.

Spectral slicing is a common mode used to increase spectral power by directing different wavelength ranges to separate spectrographs with optimized dispersers and cameras, post collimation, typically. Spectral slicing can be employed within a replicated instrument but is not a replication per se, though common optical elements can be used in the different cameras, sometimes. Replicated instruments that use spectral slicing include the SDSS spectrograph, MODS, and LAMOST. PFS, under development for Subaru, is currently an extreme example which includes three-fold spectral slicing to cover 380 – 1300 nm, and four-fold replication for field division to provide both high spectral power and grasp as seen in Figure 1.

Image slicing is the most common mode employed to enable 3-D or integral field spectroscopy, usually to increase the field of view of the observation, but also to reduce the physical size of the resolution element and increase spectral resolution, depending on the science drivers. Integral field spectrographs employ an integral field unit (IFU) that reformats the 2-D field of view for input into the spectrograph (1-D). The three primary IFU types use lenslet arrays, optical fibres (with or without input lenses), and image slicers [8, 24]. Lenslet array IFUs at the telescope focus produce micro-pupil images, which form the input to the spectrograph [25, 26, 27]. Wavelength coverage is limited to prevent overlap of spectra.

Fibres allow the field to be reformatted into a pseudo-slit input to the spectrograph [e.g. 7], and have the advantage of allowing the weight of the spectrograph(s) to be located off the telescope in a controlled environment as well as affording much greater wavelength coverage and improved packing of spectra on the detector. The disadvantages of fibres are throughput in the ultraviolet and focal ratio degradation (FRD, e.g. [28]) if used at input focal ratios slower than about f/4. Coupling fibres to the sky with microlenses, focusing a micro-pupil on the input to each fibre, allows contiguous fields to be formed and the input focal ratio to be decreased, though at the expense of coupling efficiency and spectral resolution. Bare fibres offer the most easily implemented image-slicing mode for objects distributed over wide fields of view, where single fibres of small deployable IFUs (dIFUs) can be positioned on objects of interest [29, 30, http://www.sdss3.org/future/manga.php]. The fibers can be anti-reflection (AR) coated or immersed against an AR-coated cover plate to improve coupling efficiency. As an example, VIRUS uses 75 large fixed IFUs, each with 448 fibres in a 1/3 fill-factor hexagonal close pack to achieve very large areal grasp with 33,600 fibres in total, each 1.5 arcseconds diameter on sky [31]. A "dither" pattern of three offset exposures fills in the 50 x 50 sq. arcsecond area of each IFU, enabling the huge grasp shown in Figure 1.

The more complex IFU components that are called "image slicers" come in various forms and can include both reflective and refractive micro-components. They have the advantage of maintaining spatial information along the length of micro-sliced slitlets, and hence can achieve the densest packing of spectra on the detector. This power comes with complexity and tight alignment tolerances, internally and to the telescope. The simplest versions use stacks of progressively tilted mirrors to spread the spatial elements into a stepped slit [e.g. 32]. More complex versions employ refractive elements and anamorphic effects to compress image elements and tune the field coverage and spectral resolution. MUSE uses complex field microslicers for each spectrograph channel, post field division, yielding a very high spatial sampling that is well-adapted to AO, and with high density of the spectra on the detectors [33]. The alignment and stability required for the MUSE slicer is a tour-de-force [34], but may present challenges to scale-up further. Other complex slicers have been proposed that offer the potential for high efficiency [24], but have not yet been demonstrated on sky.

On the largest telescopes, image slicing will become increasingly necessary just to sample the image sufficiently to maintain spectral resolution. This leads to more of the detector area being devoted to the spatial dimension, and provides a further driver to replication in order to maintain both wavelength and spatial coverage and object multiplex. As the spatial elements become smaller, it is also important to ensure that sufficient photons are detected from source and sky in order to overcome detector read-noise. Very-low read-noise detectors will become increasingly important even for large telescopes as the spatial elements will get smaller on sky with AO and it will become increasingly difficult to achieve background dominated observations of faint objects.

Pupil slicing is not really a mode that has been exploited, except in the wavelength dimension with spectral slicing. True pupil slicing would divide the circular pupil to different spectrograph channels, posing geometric problems that tend to negate the advantages of slicing. Combining the resulting spectra from the sliced channels would also pose data reduction problems and the reduced photon count-rate might require longer exposures in the face of detector read-noise.

**4. The first steps in large-scale replication**

Replication of spectrograph units holds the key for breaking the engineering and cost barriers reached in the current generation of instruments. If the science drivers allow, it is less expensive to make multiple copies of an instrument to gain grasp/multiplex than to build a larger monolithic instrument. Unit component costs are lower (with smaller optics and the savings of making multiple copies) and the engineering effort that typically accounts for 60-70% of the cost of a monolithic instrument is amortized over multiple copies. These advantages are not fully realised until the replication factor exceeds of order 20 and begins to asymptote around 100 copies (or "large-scale-" and "massive-replication", respectively).

The advantages of large-scale replication to increase total grasp and mitigate the size-growth of spectrographs are in the process of being embodied in the LAMOST, MUSE and VIRUS instruments. Here we discuss their properties in more detail and examine the lessons learned in realising this type of instrument for the first time, using VIRUS as an example.

In its primary mode, MUSE for the ESO Very Large Telescope (VLT) employs an advanced image slicer [33] to divide a 1 sq. arcminute Field of view into 90,000 spatial elements, each 0.2x0.2 sq. arcseconds This huge grasp is achieved by partitioning the field of view into 24 parts, which are then fed to separate integral field spectrograph units. The IFS units employ reflective slicers to divide the field into elements, which are then dispersed over a wide wavelength range of 465-930 nm at a spectral resolution R of about 3000. Each unit has refractive optics and a 4kx4k CCD detector. The total pixel count is 380 Mpxl. The spectrographs were constructed commercially and the detector system supplied by ESO [35]. MUSE is designed to work with the VLT GLAO system in order to exploit the small spatial resolution elements. MUSE is installed on the Nasmyth platform of one of the 8 m VLT telescopes and is undergoing commissioning at the time of writing. As seen in Figure 1, it promises very large grasp and spectral power, which will be used for deep surveys of distant star-forming galaxies early in the history of the universe that emit the Lyman-α line of hydrogen (LAEs).

LAMOST is a facility developed as a whole for large-scale surveys of relatively bright objects. A 4 m aperture feeds up to 4000 fibres, which are fed to 16 bench-mounted low-resolution spectrographs, each with a red and blue arm [22]. A robotic positioner sets the fibres on targets. By replicating a relatively simple spectrograph design, LAMOST achieves a high grasp, and has completed a pilot survey of 1.7 million stars and galaxies on the way to observing 5 million objects [4].

VIRUS on the 10 m HET uses replication to achieve much higher areal grasp with coarser spatial and spectral resolution [2,31]. Light gathered in the focal surface of the upgraded HET will be fed to 75 bundles of 448 fibres, each fibre subtending 1.5 arcseconds on the sky, with a 1/3 fill-factor. Each of these IFUs feeds into two identical spectrograph channels. This two-channel unit is replicated 75 times to provide 33,600 fibres in total covering 16.5 sq. arcminutes of sky, instantaneously. Light from the fibres is dispersed over 350-550 nm in catadioptric spectrographs with 125 mm beamsize, and detected on 2kx2k CCDs fed at f/1.3. The total pixel count is 630 Mpxl, binned 2x1 so very similar to MUSE, and competitive with the state-of-the-art very large DECam and HyperSuprimeCam imagers at 570 Mpxl and 870 Mpxl respectively [36,37].

HET represents an extreme among current VLTs due to its large pupil size coupled with a site that delivers 1.0 arcsec FWHM median images, exacerbating the drive to large instrument pupil-size, and making the design problem comparable to a 25 m telescope on a 0.4 arcsec site, from an instrument design point of view (see below). VIRUS is designed to survey large areas of sky rapidly, in a blind spectroscopic mode, and will be used to survey 450 sq. degrees to map a million LAEs in a 9 Gpc$^3$ volume at $1.9 < z < 3.5$ for the Hobby-Eberly Telescope Dark Energy Experiment (HETDEX [38]). These data will be used to constrain the expansion rate of the universe and the contribution of dark energy at an early epoch of the universe about 10 Gyr ago. The very large areal grasp is facilitated by the physically-large fibre core diameters, at the expense of spectral resolution. So while MUSE can survey a relatively small patch of sky to unprecedented depth, VIRUS provides shallower data over very large areas. The two instruments use large-scale replication to attack different problems and achieve similar state-of-the-art total instrument grasp and number of resolution elements, as seen in Figure 1.

The development of replicated instruments follows a different path to monolithic designs. Overall, the amount of engineering effort per unit is much lower, but there is a need to undertake more engineering effort to ensure that requirements over an ensemble of instruments are met and to speed the assembly and characterization phases. Following manufacturing best practices, a prototype must be built and preferably used on sky before freezing the design. This step can add significantly to the duration, but forces the confrontation of science requirements with instrument performance and also encourages early development of the data reduction pipeline. These factors reduce overall risk. Following any changes from the prototype, parts are ordered and the production line is set up. Before full-scale production begins, the "First Article" unit is constructed and evaluated. Ideally, the First Article is followed by a limited production run, but the scale of ~100 units that we are contemplating in astronomy does not usually necessitate this step.

In the process of developing VIRUS, we followed this protocol and learned a lot from the prototype (Mitchell Spectrograph, aka VIRUS-P [39]), which we used to undertake a pilot survey for HETDEX [40,41], ensuring the science requirements were met and software pipelines developed [42]. In taking the prototype to a replicable design, we adopted several principles and discovered rules of thumb as follows:

- Finding the lowest cost per pixel for the detector is a key driver. For VIRUS this led to the format of ~2kx2k pixels, which at the time offered very high yields and reduced risk for the manufacturer sufficiently that they were prepared to accept small margins. We purchased 200 detectors, from several wafer runs. At the present time, the balance may be starting to shift towards 4kx4k detectors, but 2kx2k format remains the lowest cost per pixel.
- Use of liquid nitrogen (LN2) to cool CCD detectors to ~-100 Celsius working temperature. Early on, we undertook a trade-off cost and reliability for the VIRUS cryogenic system which strongly favoured a distributed LN2 system over a large number of closed-cycle cryocoolers. Thermo-electric cooling is inadequate for the low-noise application of spectroscopy. VIRUS uses a novel heat exchanger to cool the detectors without having LN2 inside the cryostats, facilitating removal of the cameras for maintenance [43]. MUSE also adopted LN2 as the coolant, with a direct flow system [44].
- Following trade-off, we settled on f/1.3 as the on-axis focal ratio for the fastest camera that can be made and aligned in quantity. This focal ratio is well matched to the plate scale of 10 m class telescopes and the 15 µm pixel size of modern CCD detectors. For VIRUS we had to choose a catadioptric Schmidt-based design for both collimator and camera with a total of 6 components (including two aspheric refractive elements) due to the need for UV coverage [40,45]. The cost and throughput is very competitive with refractive designs, but necessitated a larger pupil size due to the obstruction of the detector head.
- For the optic sizes of around 200 mm used in VIRUS, the cost per component asymptotes around quantity a hundred. This is because at that point the majority of the cost is in the material, rather than the figuring.
- Volume phase holographic gratings (VPHGs) are cost-effective in quantity and offer the ability to tune the design. In large quantities, this is also true of surface-relief gratings, where a new master can be ruled, but VPHGs offer greater durability, higher efficiency, and lower scattering, particularly if the design is used in transmission, which aids certain mechanical aspects of the design. We supplied the VPHG manufacturer with a custom tester to address specifically our requirements on first order diffraction efficiency and scattered light, and over the course of production this was used to hone the production methods to improve performance [46].
- Careful design of mechanical components is needed to facilitate mass production, and the production method needs to be factored into the design. This is particularly true for small parts that were replicated in quantities up to 800 with computer numerically controlled (CNC) and electro-discharge machining (EDM) techniques [47,48]. Use of modern metrology equipment such as coordinate measuring machines (CMM) should be factored into the design for quality control and alignment.
- Use of Aluminium for all structure components, except those requiring high stability to temperature changes (e.g. focus references) for which Invar 36 was used. Flexures ensure controlled compliance between dissimilar materials. This design principle ensures alignment stability over a wide range of ambient temperature. While VIRUS is mounted in large clean-room enclosures, it sees the full range of ambient temperature change since the enclosures must not interfere with the dome-seeing environment.
- Use of castings and large monolithic parts fabricated with CNC mills. Larger mechanical components should be fit to the capacity of CNC mills and, for large parts, concentrating precision in a few features is another way to reduce overall cost [47]. On VIRUS we used aluminium and Invar 36 castings with a small amount of post machining, and even utilized cast-aluminium for the camera cryostats [47]. We did find that the time-cost of setting up castings for post-machining should be weighed against the material cost of machining a part directly out of metal stock. The outcome of the trade depends on the relative costs of machine time and labour, since it is now becoming common practice to leave machines running overnight with limited supervision.
- Simplifying assembly as far as possible and concentrating the alignment problem into a small number of components reduces the effort needed to take the spectrographs from parts to the telescope. Designing dedicated alignment tools and addressing bottle-necks by having parallel lines reduces production time, but these steps are labour intensive, still. We have found the final alignment to be a limiting step with the limited effort available. A very deterministic alignment procedure based on wavefront analysis does make it possible for less technically skilled personnel to undertake the task [49].
- VIRUS is not general purpose, so we could focus the design and choice of components in an optimized manner. Even so, the design can be adapted to a wide wavelength range as is evidenced by the new HET LRS2 low resolution spectrograph that covers 370-1050 nm by slicing the wavelength range between four VIRUS unit channels [50].

At the time of writing we are 1/3 the way through production of VIRUS and plan to begin deployment in late 2014. In total, the 75 units will contain ~15,000 parts (not including fasteners) and 700 km of fibre. All told about ½ the total cost has been expended on the spectrographs and IFUs, ¼ on the detector system, and ¼ on engineering labour and data reduction software development.

## 5. Enabling optical-mechanical technologies

The currently available technologies that enable large-scale replication and significant cost-savings are CNC grinding and polishing of optics (including aspheric surfaces, which open up design space) and CNC and EDM mass-production of mechanical parts. Some use of 3D printing with plastics for low-precision parts is coming into play. Barriers remain in the time to integrate and align multiple units, and significant engineering effort has to be applied to preparing for and executing that step.

As we look forward to a generation of replicated spectrographs beyond MUSE and VIRUS, there are several technologies on the horizon that may have important impact:
- Proliferation of deterministic CNC polishing techniques for optics will drive costs down further, though material cost will still dominate.
- Curved gratings and detectors to reduce the number of components
- Bolt-together integration of opto-mechanical assemblies, which will impact the bottle-neck of instrument integration. This is already possible but cost-prohibitive. Reflective systems can be machined from a block of aluminium, but assemblies that integrate glass optics and mechanical parts are harder.
- 3D printing of metal parts will begin to become effective for mass production, not just high-value complex parts, and this promises to have an important impact in opening up design space for monolithic parts that include features that cannot be machined, currently. As with the widespread use of CNC and EDM machines, it will take a while for designers to realise the full potential of the new technology.

- Micro-optic elements facilitate efficient coupling to the telescope and are available in sufficiently large array formats to be used as key elements within spectrographs. Arrays of micro-optics can now be built with micron-level registration between components and relative to fiducial marks used for alignment, facilitating the assembly of more complex micro-optical structures [e.g. 50].

The above discussion is currently most applicable to optical instruments, but the same principles of increasing the power of instruments through replication apply in the NIR. At the current time, however, detector cost presents a barrier to *large-scale* replication. Replication can certainly still alleviate the issue of NIR instrument cost growth since it allows for designs that are optimized to utilize the pixels as efficiently as possible and can also lead to engineering advantages for cryostat design, for example.

Extension of the large-scale replicated spectrograph concept into the NIR (with detector system costs comprising ~1/4 of the total, rather than dominating the cost) requires detector technology to become a factor of 50 cheaper on a per-pixel basis. The principles explored in MUSE and VIRUS are otherwise applicable to the NIR. Currently the only instrument that qualifies as replicated is the Subaru FMOS, with two copies of the same fibre-fed spectrograph [20]. However, they were fabricated at different institutions and are not true replicas. The prospects for a significant (tenfold) drop in detector price are poor based on the current technologies such as HgCdTe. The astronomical requirements of low noise and low dark count are not well aligned with the main market for NIR remote sensing and military night-vision applications. Inexpensive sensors based on InGaAs are available, but suffer from very high dark rate, so have limited application. It is beyond the scope of this review to speculate about where technological advances might come from, but we simply note that cost per pixel is the current barrier to large-scale replication of NIR instruments, but we expect replication to play an increasing role in NIR instruments.

## 6. Applications to the next generation of instruments for ELTs

A highly-replicated spectrograph has not been proposed as a first-light instrument for any of the ELTs. The huge grasp of MUSE and VIRUS makes them quite adaptable to ELTs and certainly the scale of these instruments is "extremely large" (MUSE takes up a whole Nasmyth platform at VLT and VIRUS fills two clean-room enclosures, each 6x6x1.5 m^3).

Three ELTs are under development. All have a significant science emphasis on spatial resolution supported by AO, but each has a wide field survey spectrograph designed to work with natural seeing or GLAO. In Table 3 and Figure 2 we present the grasp and spectral power of proposed ELT instruments DIORAMAS (EELT[6]) WFOS/MOBIE (TMT [51,52]) and GMACS (GMT [5]), compared to VIRUS and MUSE and other replicated instruments on VLTs. DIORAMAS has 4 replicated channels and GMACS is plotted for a four-spectrograph instrument that will probably strain budgets. The locations of the ELT instruments indicate why high grasp through replication will become essential to exploit the vast amount of information being gathered, not just for large surveys as is the focus on smaller telescopes.

As an example of how large-scale replication can address some of the challenges of instrumenting the next generation of telescopes, we consider how VIRUS might be adapted to the 25 m Giant Magellan Telescope. As mentioned above, the relatively poor image quality of HET coupled with its 10 m pupil size qualify HET as the world's smallest ELT, since the instrument challenges are very similar. This is trivially illustrated by the fact that the VIRUS fibers project to 1.5 arcseconds at HET and would subtend 0.6 arcseconds on GMT, well matched to the excellent site seeing at the Las Campanas site. In principle, VIRUS could observe 16,000 objects in traditional single-fibre MOS mode, with a sky fibre assigned to each object, to ensure adequate sky-subtraction on faint targets, though the fibre positioning system would be frightening.

More interesting would be to couple VIRUS with an adaptive focal surface system such as the proposed Many Instrument Fiber System (MANIFEST [53]) and use the grasp of VIRUS to feed dIFUs of a range of sizes. VIRUS units could also be configured with different wavelength ranges and resolutions depending on the application, and no doubt the balance of detector area would be shifted towards wavelength coverage and resolving power. This configuration is very similar to that proposed for GMACS in combination with MANIFEST. The 4-unit GMACS would accommodate 420 dIFUs, and cover a wide spectral range at higher resolution than the lower resolution imaging spectrograph mode. The 0.25 arcsecond diameter spatial element of the dIFU is much better matched to the camera focal ratio and pixel size of GMACS, which is why the GMACS/MANIFEST point stands out as a maximum in TIG in Figure 2. This point illustrates the power of image slicing for ELT applications, as discussed in Section 3, since the slicing allows much better use to be made of the total pixel count in the large CCD arrays proposed for GMACS.

As a specific example of what VIRUS could do in this context, it would be possible to deploy 2000 19-element dIFUs with 1.7 arcsec field, sampled at 0.33 arcseconds, accommodated by VIRUS with 150 μm diameter fibers. These dIFUs are very similar to those considered for the GMACS/MANIFEST combination. The MANIFEST positioning system could deploy these 2000 dIFUs anywhere in the 20 arcminute field of GMT and the wavelength ranges of the VIRUS units could be selected by object. The resulting VIRUS point is plotted on Figure 2 as VIRUS/MANIFEST. Note that this highly competitive replicated spectrograph is identical to VIRUS except for an exchange of dispersers and only uses 2kx2k CCDs. It has not been optimized for the GMT, other than to consider the configuration of the IFUs. Based on experience with VIRUS, it could be realized for ~$20M, a fraction of the projected cost for GMACS or the other ELT instruments.

This toy example simply illustrates the power that replication can bring to bear on the challenges posed by the 50 million spatial resolution elements available in the GMT or other ELT field of view, even without any AO. Of order a percent of the spatial elements will have an object, many of which will be of interest, so taking full advantage of these telescopes will be a significant challenge that replication can begin to address. Thousands of targets will be available, rather than hundreds. The ability to feed multiple instruments, whether VIRUS-like or not, through an adaptable targeting system like MANIFEST will also be an important part of any solution.

## 7. The photonic crystal ball

The above discussion addresses replication of spectrographs of a more traditional design, where the instrument employs physical optics and is working, typically, far from the diffraction limit. The next frontier is the integration of photonic structures into replicated spectrographs. This technology has only just begun to show up in astronomical applications, but the prospect of writing fully-photonic spectrograph structures has the potential to make true mass-production a reality.

Advances in photonics technology hold potential for escaping from the physical optics regime to realise truly replicated modular spectrographs, and the reader is referred to reviews on the applications of photonics to astronomy [54,55,56] for more details. These technologies have been incorporated into the photonic integrated multimode microspectrograph (PIMMS) concept [55]. The first technology advance needed to create an integrated photonic spectrograph (IPS) is the photonic lantern (PL) which sorts a multimode input into single mode (SM) fibers or waveguides with high efficiency [e.g. 57]. The SM fibers/waveguides can be arrayed in a line to form a pseudo-slit for input to a spectrograph. Hence, in its simplest form, the PL allows a telescope to be coupled to a diffraction-limited spectrograph with high efficiency, after which the performance of the spectrograph is independent of the telescope [55,58]. Even with AO on ELTs, the inputs will not be 100% diffraction-limited and will require a multi-mode input [59].

The second technology for a true IPS is array waveguide dispersers, which can provide efficient dispersion of light in a much smaller volume than diffraction gratings [55,60,61,62,63]. Precise phase shifts, achieved through path differences in an array of waveguides, lead to a dispersed spectrum with overlapping orders when the light free-propagates at the output of the array waveguide. Individual orders can be isolated by blocking filters or separated with cross-dispersion [55,60]. If the SM inputs to the array waveguide are spaced such that their free spectral ranges do not overlap at the output, dubbed a cyclic array waveguide [55,60,62,63], the multiplex of the system can be increased, though there are limitations and real-world examples are being developed [60,63].

High grasp would still require a large number of pixels. A multimode input must first be separated into SM waveguides with a photonic lantern. The number of SM guides depends on the image quality delivered by the telescope. GLAO will help this factor over wide fields, but there is still a multiplier of order ten, or more. The number of AWGs and number of detector pixels are multiplied accordingly. Whether the required arrays of AWGs will be a cost-effective replacement for diffraction gratings remains to be seen. Once in the diffraction-limited regime, however, the scale of individual channels is independent of the telescope aperture, and replication can be used to scale the instrument for the particular telescope or application.

Additional photonic refinements such as the introduction of Bragg gratings (BGs) in the SM fibers or waveguides, after the photonic lantern input, can also enable powerful modes such as selective suppression of OH emission lines from the sky at red and NIR wavelengths [64,65]. At the time of writing, fully integrated photonic spectrographs and OH suppression have been demonstrated on sky [66,67], with promising results, but several more years of development will be needed before these technologies can be considered mature enough to form the basis for competitive astronomical instruments. Fortunately the main driver for photonics devices comes from the telecommunications industry, and astronomy can benefit from the huge investment (as it has from the development of efficient fibre-optics, also a photonic technology).

The final necessary technology for an IPS is energy detection. As mentioned above, the cost per pixel is paramount in making large-scale replication affordable. Most of the telecom industry investment in photonics is for NIR wavelengths, where the lack of affordable detectors would hold back replication (see above). The added driver for diffraction-limited spectrographs is the need for small pixels, on the order of 5 μm, coupled with fast cameras to utilize the pixels efficiently. Detector read noise is an issue with small pixels, as efficient observations must be photon- rather than read-noise dominated. Pixels with rectangular format can aid in this [55]. The ideal detector for an IPS would have energy resolution so as to remove the need to isolate orders through blocking or cross dispersion, thereby maximizing the information gathered. It would also be a photon-counting detector with zero read noise, since background levels for faint objects may be very faint with small pixels and OH background suppression. Finally, it would have broad wavelength coverage extending into the NIR.

While no detectors in common use fulfil these ideals, microwave kinetic inductance detectors (MKIDs [68,69,70]) hold significant promise. MKIDs operate at superconducting temperatures and can measure the energy and arrival time of single photons, using microwave multiplexing to access individual pixels tuned to specific frequencies. The energy resolution of MKIDs (R<100) is not sufficient for most astronomical spectroscopy, but in combination with photonic technologies, they can effectively provide the cross-dispersion needed to separate orders from cyclic array waveguides without resorting to physical optics. MKIDs are a very new technology, but a 2024 pixel array has been used for astronomy [70], and they are much more scalable to large arrays than other superconducting detectors with energy sensitivity (superconducting tunnel junctions [71] and transition edge sensors [72]). An MKID array, mated to an IPS could eventually cover 300 to 1800 nm at R~1000, simultaneously (i.e. from the UV atmospheric cut-off to where the thermal background dominates), and be replicable on sufficiently large scale to form the practically ideal astronomical survey spectrograph.

Certainly photonic components such as PLs and BGs will be a feature of future replicated spectrographs. We are some way from realising the full combination of these technologies into a fully-photonic spectrograph with integrated detector, even for a single instance, though the PIMMS instrument is an important step [55]. The biggest challenge is likely to be achieving high efficiency along with high grasp. However, once an integrated photonic spectrograph can be produced, the prospect of a truly replicated spectrograph able to deal with the information gathered by an ELT, and even able to reject unwanted contaminating light from the sky, is much closer. Such an instrument also promises increased multiplex in a given instrument volume [63].

## 8. Summary

Since the 1980s the advent of large format, highly efficient, low-noise CCD and HgCdTe detector arrays for optical and NIR wavelengths has led to a revolution of capability for astronomical spectrographs, especially in the area of wide area surveys of large samples of objects. Manufacturing techniques for optics and gratings have also achieved high efficiency. The scalability of this model of instrument design is now hitting a ceiling where the size of detectors and optics and the difficulty of engineering spectrographs are reaching a limit. Pushing beyond, as required for the next generation of ELTs, is going to need a change in paradigm, and replication appears to offer a cost-effective way to instrument these telescopes.

The overriding principle in replicated spectrographs is that the detector area be utilized efficiently, as with conventional monolithic instruments. This implies fast camera optics and the lowest possible cost per detector pixel. A key advantage of replication other than economies of scale is that the physical size of the units can be kept small enough that standard manufacturing techniques can be applied to optics and other components. The main downside is that instrument volume increases and the cryogenic systems required to cool the detectors become more complicated, due to their distributed nature.

VIRUS and MUSE offer two fairly different worked examples of large-scale replication of spectrograph channels, and promise to be highly effective when they come on line in the next year. They are both integral field spectrographs, and while they have similar TIG, they have quite different focus – MUSE goes very deep with high spatial resolution over a limited field of view, while VIRUS will enable very wide-angle blind spectroscopic surveys for the first time, albeit at a shallower level.

For the time being, replication of moderately-sized spectrographs of more conventional design offers the best prospect for instrumenting ELTs. New photonic and detector technologies hold significant promise, but will take time to mature to a competitive level. In the future, the key for replication to be effective will be to match the extraordinary efficiency that has been achieved for monolithic spectrographs, as we move into a regime where replication on an industrial-scale is going to be our next tool to gather the maximal information from a unit exposure on an ELT.

**Acknowledgments**
Steve Rawlings provided much encouragement and motivation for VIRUS and the idea of large-scale replication. Fancisco Cobos provided much insight on optical design. This review is dedicated to their memories. I am particularly grateful to Matthew Bershady for sharing his compilation of instrument properties and for motivating the discussion of grasp as a metric, as well as for many insightful discussions. Joss Bland-Hawthorn and Roger Haynes provided much insight on the current state of astro-photonics, as did Michael Lesser and Christoph Kutter on detector technology evolution. The reports of two anonymous referees helped clarify important points.

VIRUS is led by the University of Texas at Austin McDonald Observatory and Department of Astronomy and is a collaboration with the Ludwig-Maximilians-Universität München, Max-Planck-Institut für Extraterrestriche-Physik (MPE), Leibniz-Institut für Astrophysik Potsdam (AIP), Texas A&M University, Pennsylvania State University, Institut für Astrophysik Göttingen, University of Oxford and Max-Planck-Institut für Astrophysik (MPA).  In addition to Institutional support, VIRUS is funded by the National Science Foundation (grant AST-0926815), the State of Texas, the US Air Force (AFRL FA9451-04-2-0355), by the Texas Norman Hackerman Advanced Research Program under grants 003658-0005-2006 and 003658-0295-2007, and by generous support from private individuals and foundations.

VIRUS is the result of nearly a decade of development and particular thanks go to the following for their leading contributions to development of the instrument: Joshua Adams,

Darren DePoy, Niv Drory, Maximilian Fabricius, Karl Gebhardt, Dionne Haynes, Thomas Jahn, Andreas Kelz, Hanshin Lee, Phillip MacQueen, Jennifer Marshall, Jeremy Murphy, Joe Tufts, Sarah Tuttle, and Brian Vattiat. Thanks also to the following for their important contributions to motivating, designing, funding, and building VIRUS: Aditi Raye Allen, Richard Allen, Heiko Anwad, Sam Barden, Joel Barna, Frank Bash, Svend Bauer, Ralf Bender, Guillermo Blanc, Emily Booth, John Booth, Emily Boster, Chris Cabral, Taylor Chonis, Chris Clemens, Amanda Collins, Mark Cornell, Gavin Dalton, Michael Denke, Eric Dreasher, John Good, Frank Grupp, Marco Haeuser, Ulrich Hopp, Donghui Jeong, Ingrid Johnson, Eiichiro Komatsu, Herman Kriel, David Lambert, Martin Landriau, Bob Leach, Kayla Leonard, Mike Lesser, Francesco Montesano, Harald Nicklas, Povilas Palunas, Dave Perry, Andrew Peterson, Trent Peterson, Travis Prochaska, Mary Ann Rankin, Marc Rafal, Marisela Rodriguez, Martin Roth, Richard Savage, Don Schneider, Mike Smith, Jan Snigula, Matthias Steinmetz, Carlos Tejada, and Gordon Wesley.

Vendors of major components for VIRUS include: Asphericon, Astronomical Research Cameras, Cascade Optical, CeramOptec, Corning Tropel, Cosmo Optics, Fiberware, Harold Johnson Optical Labs, Leoni Fibertech, Midwest Cryogenics, MKS, PG&O, SyZyGy, and University of Arizona Imaging Technology Lab.

We thank the following for acting as reviewers of VIRUS: Roland Bacon, Gary Bernstein, Gerry Gilmore, Rocky Kolb, Richard Kurz, Adrian Russell, and Ray Sharples. We thank the staffs of McDonald Observatory, AIP, MPE, Texas A&M, IAG, and Oxford University Department of Physics, for their contributions to the development of VIRUS.

Table 1 - High Grasp Spectrographs of Monolithic Design

| Instrument | Type | Telescope (aperture) | Total sky area (sq. arcmin) | Areal element (sq. arcsec.)* | Number areal elements | Resolving power | Number of Spectral elements | Reference |
|---|---|---|---|---|---|---|---|---|
| IMACS | MOS/slitlet | Magellan (6.5 m) | 1.08 | 3.00 | 1296 | 600 | 556 | 12 |
| AAOmega | MOS/fiber | AAO (4 m) | 0.38 | 3.46 | 392 | 1300 | 580 | 13 |
| KMOS | NIR/advanced-slicer | VLT (8 m) | 0.05 | 0.04 | 4204 | 3600 | 1000 | 14 |
| HERMES | MOS/fiber | AAO (4 m) | 0.37 | 3.14 | 392 | 28000 | 4400 | 11 |
| WEAVE | MOS/fiber | WHT(4 m) | 0.37 | 1.33 | 1000 | 5500 | 5000 | 9 |
| 4MOST | MOS/fiber | VISTA (4 m) | 0.74 | 1.77 | 1500 | 5000 | 3900 | 10 |

* for slitlet spectrographs the areal element is the slit width multiplied by 5 arcsec, unless a specific multislit mode is described

Table 2 - Replicated Spectrographs

| Instrument | Type | Telescope (aperture) | Number of Spectrographs | Total sky area (sq. arcmin) | areal element (sq. arcsec.) | Resolving power | Number of Spectral elements | Reference |
|---|---|---|---|---|---|---|---|---|
| 4-shooter | slit/sltless MOS | Hale (5 m) | 4 | 0.22 | 1.50 | 660 | 360 | 15 |
| 2dF | MOS/fiber | AAO (4 m) | 2 | 0.38 | 3.46 | 880 | 490 | 16 |
| VIMOS | MOS/slitlet/IFU | VLT (8 m) | 4 | 0.20 | 0.45 | 2500 | 500 | 17 |
| DEIMOS | MOS/slitlet | Keck (10 m) | 2 | 0.20 | 0.75 | 2620 | 1680 | 18 |
| SDSS/BOSS | MOS/fiber | APO/SDSS (2.5 m) | 2 | 0.87 | 3.14 | 1500 | 2800 | 19 |
| FMOS | NIR MOS/fiber | Subaru (8.3 m) | 2 | 0.13 | 1.13 | 600 | 900 | 20 |
| MODS | MOS/slitlet | LBT (8.4 m x 2) | 2 | 0.12 | 0.60 | 2000 | 2760 | 21 |
| LAMOST | MOS/fiber | LAMOST (4 m) | 16 | 1.96 | 1.77 | 1800 | 1500 | 22 |
| MUSE | IFS/slicer | VLT (8 m) | 24 | 1.00 | 0.04 | 3000 | 2000 | 1 |
| VIRUS | IFS/fiber | HET (10 m) | 150 | 16.5 | 1.77 | 700 | 330 | 2 |
| PFS | MOS/fiber | Subaru (8.3 m) | 4 | 0.7 | 1.00 | 3000 | 1300 | 23 |

Table 3 - Proposed high-Grasp ELT Spectrographs Compared to VIRUS

| Instrument | Type | Telescope (aperture) | Number of Spectrographs | Total sky area (sq. arcmin) | areal element (sq. arcsec.) | Resolving power | Number of Spectral elements | Reference |
|---|---|---|---|---|---|---|---|---|
| DIORAMAS | slitlet | EELT (40 m) | 4 | 0.33 | 2.5 | 300 | 1200 | 6 |
| WFOS/MOBIE | slitlet | TMT (30 m) | 1 | 0.12 | 3.75 | 1000 | 1333 | 51, 52 |
| GMACS | slitlet | GMT (25 m) | 4 | 0.42 | 0.7 | 2000 | 1600 | 5 |
| GMACS/MANIFEST | fiber dIFU | GMT (25 m) | 4 | 0.14 | 0.05 | 5600 | 4480 | 53 |
| VIRUS/MANIFEST | fiber dIFU | GMT (25 m) | 150 | 1.04 | 0.09 | 1330 | 585 | 2 |

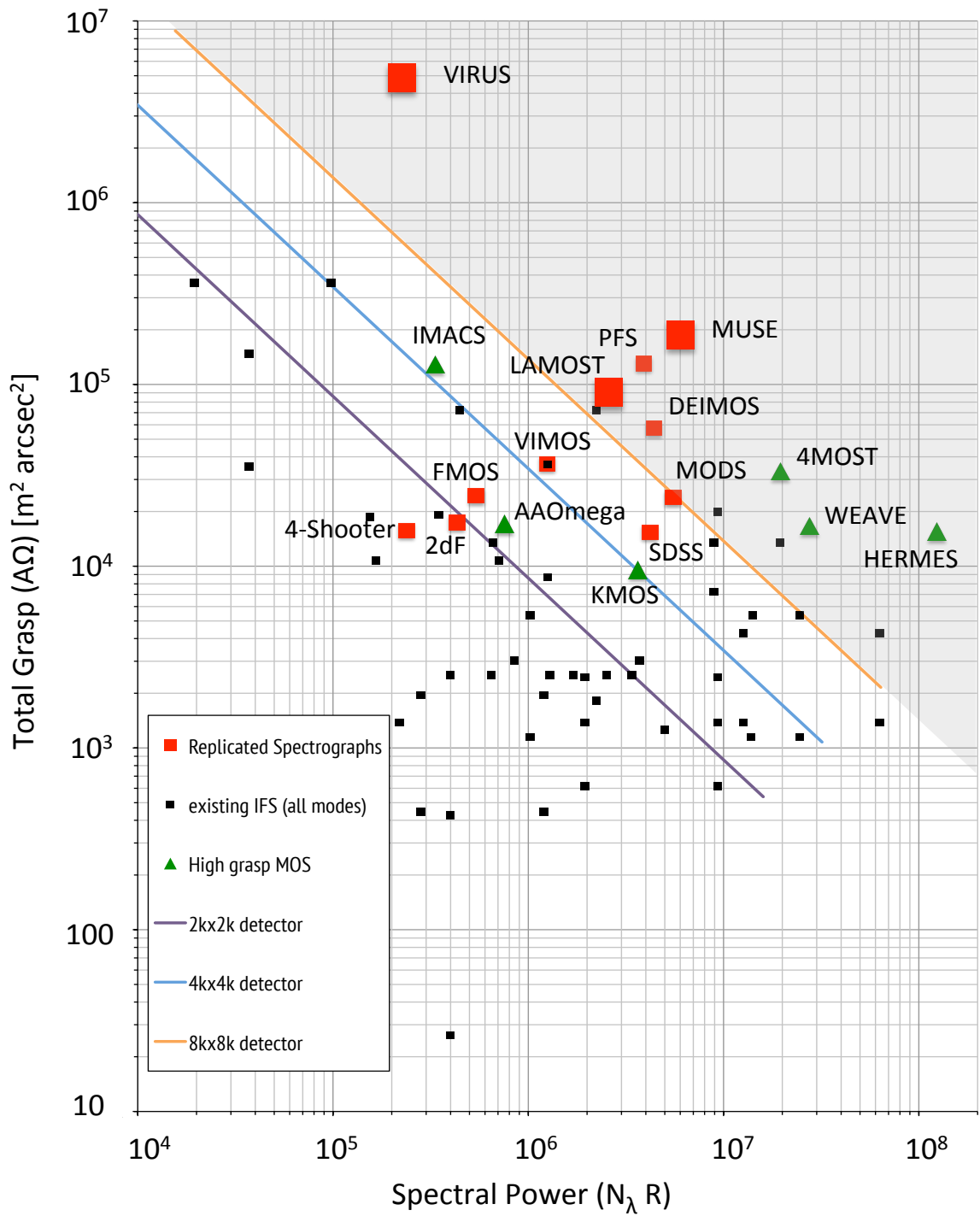

Figure 1 – Total Grasp vs. Spectral Power for replicated spectrographs (red squares), compared to integral field spectrographs (IFS, black dots) and high-grasp multi-object spectrographs (MOS, green triangles). IFS points show all modes of each instrument. See Tables 1 and 2 for details. The three instruments with more than 10-fold replication are shown by larger symbols. Diagonal lines show the loci of detectors of different total pixel count, increasing by a factor of 4 between each line. The grey shaded area is the region of the graph where detector formats larger than 8kx8k pixels are required, which is argued to be the domain of replicated instruments. Individual instruments are labeled, except for IFS. All instruments are optical except KMOS and FMOS. See text for details.

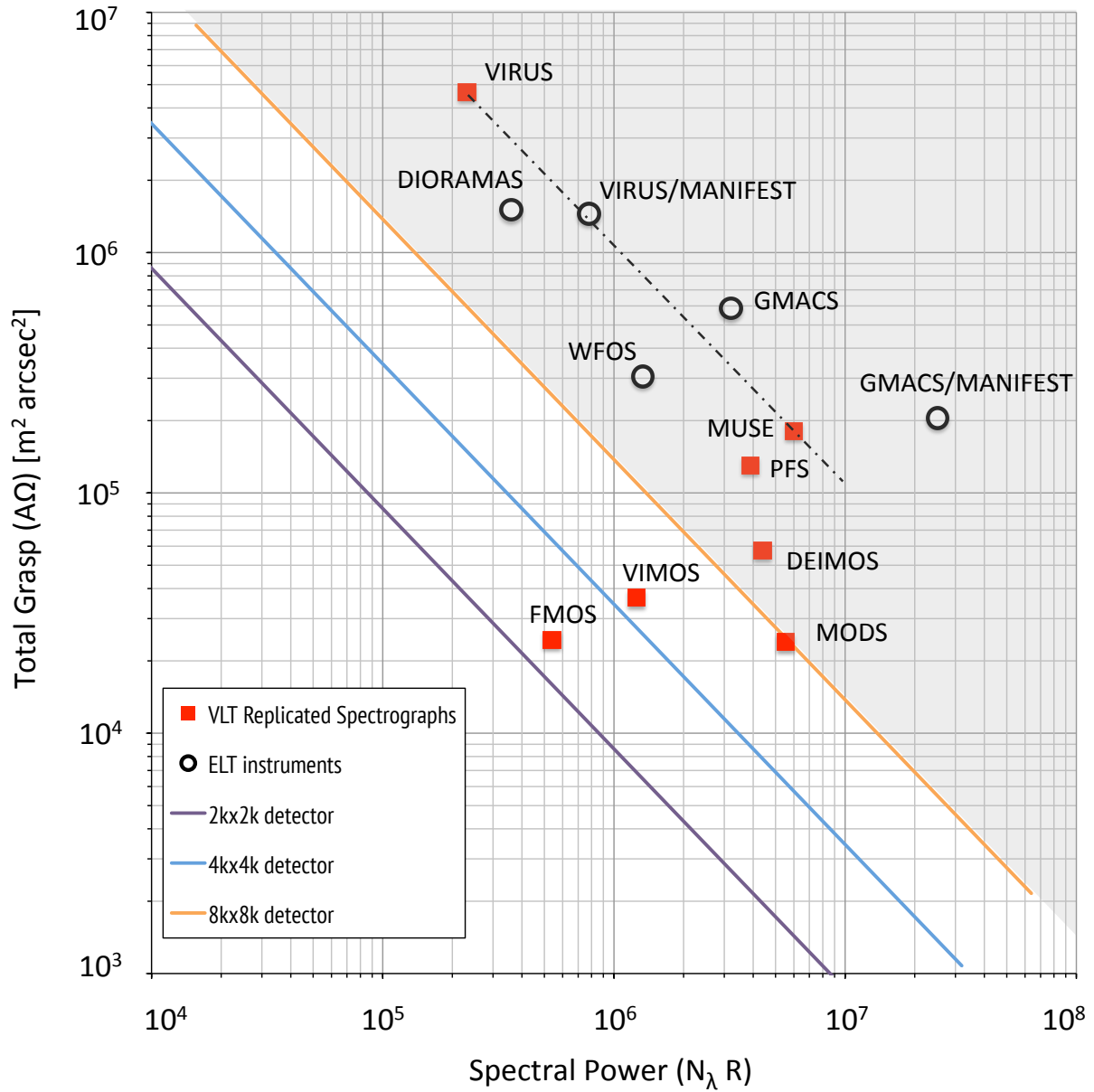

Figure 2 – Total Grasp vs. Spectral Power for replicated spectrographs on VLTs (red squares), compared to proposed high-grasp spectrographs for ELTs (black open circles). Diagonal lines and shading as in Figure 1. The dashed black line shows the locus of total grasp for a VIRUS- or MUSE-like instrument. By trading spatial resolution elements for spectral resolution elements, the replicated spectrographs can be moved along this locus. Individual instruments are labeled. See text for details.